\documentclass{emulateapj}

\usepackage{amsmath}
\usepackage{graphicx}
\usepackage{natbib}
\usepackage{bm}

\newcommand{\hii}{H~{\sc ii}}
\newcommand{\hi}{H~{\sc i}}

\def\farcs{\hbox{$.\!\!^{\prime\prime}$}}

\begin{document}

\title{
The Interplay of Magnetic Fields, Fragmentation and Ionization Feedback in High-Mass Star Formation
}

\author{Thomas Peters\altaffilmark{1,2}, Robi Banerjee\altaffilmark{2}, Ralf S. Klessen\altaffilmark{2} and Mordecai-Mark Mac Low\altaffilmark{3}}
\email{thomas.peters@ita.uni-heidelberg.de}

\altaffiltext{1}{Fellow of the Baden-W\"{u}rttemberg Stiftung}
\altaffiltext{2}{Zentrum f\"{u}r Astronomie der Universit\"{a}t Heidelberg,
Institut f\"{u}r Theoretische Astrophysik, Albert-Ueberle-Str. 2,
D-69120 Heidelberg, Germany}
\altaffiltext{3}{Department of Astrophysics, American Museum of Natural History, 
79th Street at Central Park West, New York, New York 10024-5192, USA}

\begin{abstract}
Massive stars disproportionately influence their surroundings.  How
they form has only started to become clear recently through radiation
gas dynamical simulations.  However, until now, no simulation has
simultaneously included both magnetic fields and ionizing radiation.
Here we present the results from the first radiation-magnetohydrodynamical
(RMHD) simulation including ionization feedback, comparing an RMHD model of a 
1000~$M_{\odot}$ rotating cloud to
earlier radiation gas dynamical models with the same initial density
and velocity distributions.  We find that despite starting with a
strongly supercritical mass to flux ratio, the magnetic field has
three effects.  First, the field offers locally support against gravitational
collapse in the accretion flow, substantially reducing the amount of
secondary fragmentation in comparison to the gas dynamical case.
Second, the field drains angular momentum from the collapsing gas, further increasing
the amount of material available for accretion by the central,
massive, protostar, and thus increasing its final mass by about 50\% from
the purely gas dynamical case.  Third, the field is wound up by the rotation
of the flow, driving a tower flow.  However, this flow never achieves
the strength seen in low-mass star formation simulations for two
reasons: gravitational fragmentation disrupts the circular flow in the
central regions where the protostars form, and the expanding \hii\
regions tend to further disrupt the field geometry.  
   Therefore,
ionizing radiation is likely to dominate outflow dynamics in 
regions of massive star formation.
\end{abstract}

\maketitle
\section{Introduction}
\label{sec:intro}

Massive stars influence their surroundings through
their ionizing radiation, winds, and supernova explosions.  However,
their formation remains less well understood than that of low mass
stars \citep{maclow04}. Massive stars accrete much faster than low mass
stars at rates reaching as high as $10^{-3} M_{\odot} \mathrm{ yr}^{-1}$
\citep[e.g.][]{beutheretal02}. This is required to build up the mass of the
star prior to exhausting its nuclear fuel \citep{ketoetal06}. They are
observed to have companions with far greater frequency than low mass
stars \citep{lada06,zinnyork07}.  The ionizing radiation they emit produces a
set of observables distinctly different from low mass star formation. 

Magnetic fields pervade neutral atomic gas in the interstellar medium at strengths high
enough to prevent isotropic gravitational collapse
\citep{heiles76}.  However, gravitational collapse along field lines can
proceed unimpeded, and, in some circumstances, magnetic fields can
even accelerate collapse through magnetic braking \citep{kimetal03,hencia09}.
Observations of low-mass star forming regions show that
magnetic field strength remains roughly constant at values of $B = $ 5--10
$\mu$G at number densities $n < 10^3\,$cm$^{-3}$, with the strength
growing as $B \propto n^{2/3}$ at higher number densities \citep{crutcher09}.

The role of magnetic fields in low mass star formation has received
extensive attention.  They act in three major ways.  First, they can
provide support against gravitational collapse.  Second, they drain
angular momentum directly through magnetic braking, as well as
indirectly through magnetorotational instability driven turbulent
viscosity.  Finally, as one consequence of angular momentum transport,
they drive jets and outflows.

The critical mass-to-flux ratio at which magnetic fields
can no longer support a gas cloud against gravitational collapse is \citep{mouspi76,mou91} 
\begin{equation}
\left(\frac{M}{\Phi}\right)_{\mathrm{cr}} = \frac{\zeta}{3
  \pi}\left(\frac{5}{G}\right)^{1/2} = 490 \mbox{ g G}^{-1} \mathrm{
  cm}^{-2}
\end{equation}
where $\zeta = 0.53$ is the value for a uniform sphere
\citep{stritt66}. For a field of 10~$\mu$G, this corresponds to
a column density of $5 \times 10^{-2}$~g~cm$^{-2}$, well below the
0.1-1~g~cm$^{-2}$ characteristic of massive star forming regions
\citep[e.g.][]{muelleretal02,mckeetan03,krummmcke08}. As a result, less attention has
been paid to the dynamics of magnetic fields in massive star forming
regions where it appeared unlikely that they would prevent star
formation (see, however, Hennebelle et al. 2010, A\& A submitted).

However, even in models of low-mass star formation starting with
supercritical values of $(M/\Phi)$, fragmentation is
reduced compared to the pure gas dynamical case by a combination of magnetic
pressure and tension forces providing additional support to
counteract gravitational collapse
\citep{ziegler05,banerjee06b,pribat07,henfro08,hentey08,hencia09,cometal10,buerzleetal10}.

Magnetic fields can also drain angular momentum from rotating
clumps and cores even if they do not entirely prevent collapse.
Magnetic braking in idealized perpendicular and parallel cases was
computed by \citet{moupal79,moupal80}, who found the criterion for
braking to be effective is that outgoing helical Alfv\'en waves pass
through a mass of external gas equal to that of the cloud
\citep[see also][]{basmou94,hencia09}.

Magnetorotational instability further drives outward angular momentum
transport within differentially rotating disks, or indeed any
differentially rotating structure with angular velocity decreasing
outward \citep{balbushawley91,balbushawley98}.  This may be less
important in massive star forming regions where gravitational
instability dominates angular momentum transport \citep{petersetal10a,petersetal10c}.

Outflows have been found almost universally around forming low mass
stars.  Controversy persists about the exact mechanism driving them
\citep{ferretal06}, with viable suggestions including
disk winds \citep{blandfordpayne82,koniglpudritz00}, tower flows
produced by wound up toroidal field
\citep{tomisaka98,tomisaka02,matstomi04,machidaetal04,banerjee06b,banerjee07}
or interactions between the disk field and protostellar
magnetosphere, either at an X-point \citep{najitashu94,shuetal94,shuetal07,caietal08}
or in more general geometries
\citep{lovelaceetal99,romanovaetal09}.  However, all these
mechanisms depend on the interaction of magnetic fields with
a coherent and well-defined rotational flow in the accretion disk.

In this work we use numerical simulations to examine more carefully the dynamics of magnetic fields
during the formation of massive stars, focussing on the clump scale of
0.1 to 0.001 pc. We model the collapse of a magnetized, rotating clump
containing 1000~$M_{\odot}$.  This work is complementary to the
initial numerical study of magnetic fields in massive star forming regions by
\citet{banerjee07} in that it examines a larger mass region at
larger scales. It is also complementary to recent radiation-magnetohydrodynamical
simulations of low-mass star formation by \citet{tomidaetal10} because
we take the feedback by ionizing and non-ionizing radiation into account.
We show that the three ways that magnetic fields act
in low mass star formation recur in massive star formation, but with
different implications and relative importance.

In Section~\ref{sec:method} we describe our numerical simulations,
while in Section~\ref{sec:results} we discuss our results.  Finally,
we summarize and conclude in Section~\ref{sec:summary}. 

\section{Numerical Method and Initial Conditions}
\label{sec:method}

We present the first three-dimensional, radiation-magnetohydrodynamical simulations of
massive star formation, taking into account heating by both ionizing and non-ionizing radiation,
using the adaptive-mesh code FLASH \citep{fryxell00}. We propagate the radiation on the adaptive
mesh with our extended version of the hybrid characteristics raytracing method \citep{rijk06,petersetal10a}.
We use sink particles \citep{federrathetal10} to model young
stars. Sink particles are inserted when the Jeans
length of collapsing gas can no longer be resolved on the adaptive
mesh.  They continue to accrete any high-density
gas lying within their accretion radius. We use the sink particle mass and accretion rate
to determine the radiation feedback with a prestellar model
\citep{petersetal10a}. The radiation-magnetohydrodynamical equations are solved
with a novel, positive-definite, MUSCL-Hancock, Riemann solver \citep{waagan09}.

In this work we present simulations that incorporate magnetic fields
into the gas dynamical collapse models presented in
\citet{petersetal10a}.  We further analyzed those models with a focus on the accretion
history of the stellar cluster \citep{petersetal10c}, as well as the resulting
\hii\ region morphologies \citep{petersetal10b}.
The initial conditions for the gas and the simulation parameters in
the simulations presented here match those in our previous
simulations. We start with a $1000\,M_\odot$ molecular cloud having a constant density core
with $\rho = 1.27 \times 10^{-20}\,$ g\,cm$^{-3}$ within a radius of $r = 0.5\,$pc, surrounded by
an $r^{-3 / 2}$ density fall-off out to $r = 1.6\,$pc.
The cloud rotates as a solid body with an angular velocity
$\omega = 1.5 \times 10^{-14}\,$s$^{-1}$. The initial temperature is $T = 30\,$K.
The highest resolution cells on our adaptive mesh have a size of $98\,$AU. Sink particles are
inserted at a cut-off density of $\rho_{\mathrm{crit}} = 7 \times 10^{-16}\,$g\,cm$^{-3}$ and
have an accretion radius of $r_{\mathrm{sink}} = 590\,$AU.

We compare the results of four different simulations (see
Table~\ref{tab:colsim}), three of which were discussed and analyzed in
our previous work
\citep{petersetal10a,petersetal10b,petersetal10c}. In the first
simulation (run~A), a dynamical temperature floor is
introduced to suppress secondary fragmentation. Only one sink particle
(representing a massive protostar) is allowed to form. In the second
simulation (run~B), secondary fragmentation is allowed, and many sink
particles form, representing a group of stars, each contributing to the radiative
feedback. The third
simulation (run~D) is a control run in which secondary fragmentation
is still allowed, but no radiation feedback is included. 

The new, fourth simulation (run~E) is a magnetized version of run~B,
the full stellar group simulation with radiation feedback from all
stars. Run~E includes an initially homogeneous magnetic field along
the rotation axis of the cloud with a magnitude of $10\,\mu$G,
corresponding to $(M/\Phi) = 14 (M/\Phi)_{\mathrm{cr}}$ in the central
core at the beginning of the simulation
and a plasma beta ($\beta_\mathrm{pl} = p_\mathrm{th} / p_\mathrm{mag}$
with the thermal and magnetic pressure $p_\mathrm{th}$
and $p_\mathrm{mag}$, respectively) of $\beta_\mathrm{pl} = 3.7$
. This strength was chosen to
be consistent with an analysis of combined \hi\,, OH, and CN Zeeman
measurements that yielded a field strength
\begin{equation}
B \simeq (10\,\mu\,\mathrm{G}) \left(\frac{n}{300 \mathrm{\,cm}^{-3}}\right)^{0.68}
\end{equation}
(Crutcher et al. 2010, in prep; \citealt{crutcher10}). 
Our choice of a uniform field initially neglects the increase
of field strength with increasing density, but this occurs quickly
as the collapse proceeds.
   We note that CN Zeeman observations by \citet{falgaroneetal08}
   suggest a lower mass-to-flux ratio for high-density regions, as we
   would expect.
\begin{table}
\begin{centering}
\caption{Overview of collapse simulations. \label{tab:colsim}}
\begin{tabular}{cccc}
\tableline
\tableline
Name & Radiative Feedback & Multiple Sinks & Magnetic Fields\\
\tableline
Run A & yes & no  & no\\
Run B & yes & yes & no\\
Run D & no  & yes & no\\
Run E & yes & yes & yes\\
\tableline\\
\end{tabular}
\end{centering}
\end{table}

\section{Results and Discussion}
\label{sec:results}

\subsection{Accretion History}
\label{sec:accrhist}

\begin{figure}
\centerline{\includegraphics[height=170pt]{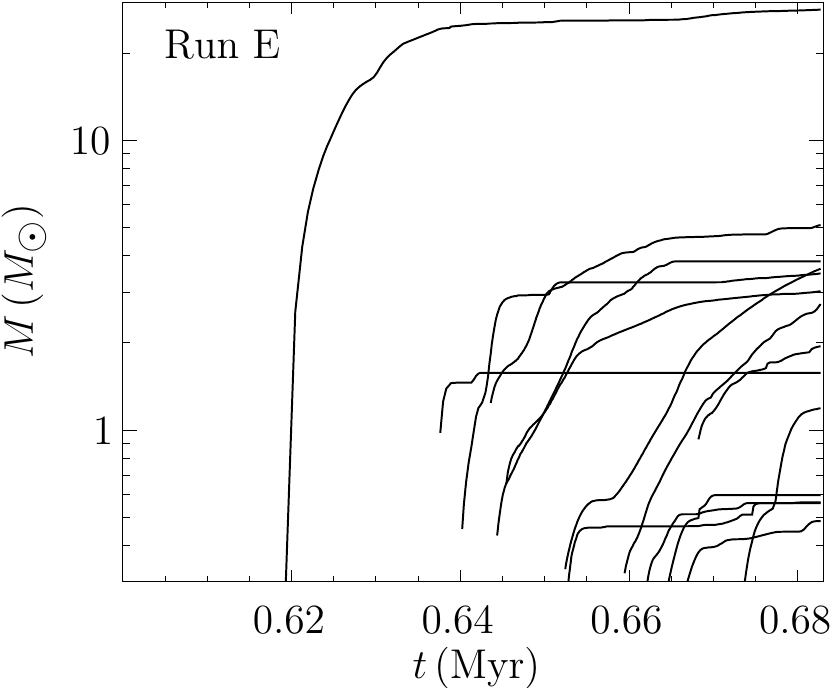}}
\centerline{\includegraphics[height=170pt]{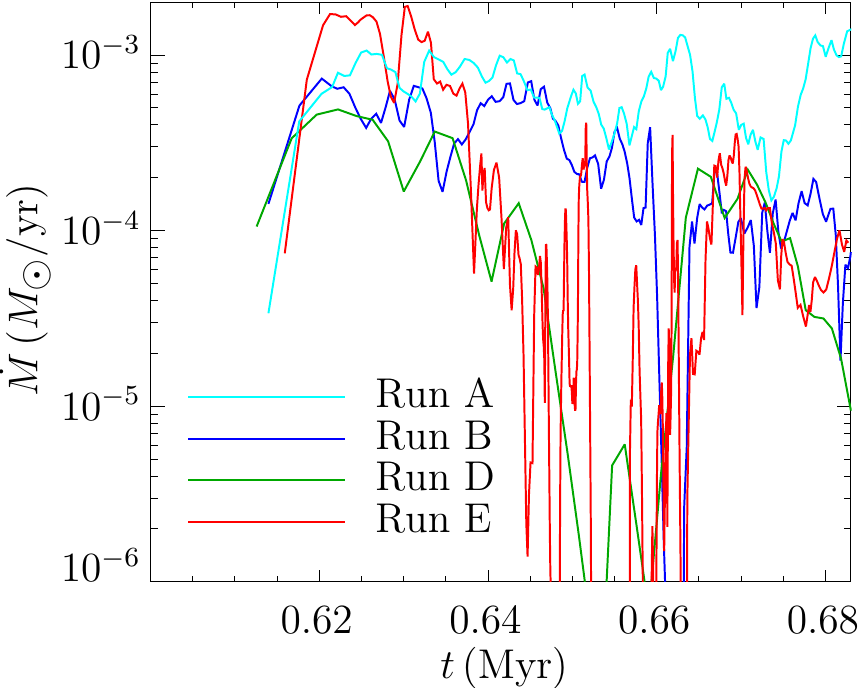}}
\centerline{\includegraphics[height=170pt]{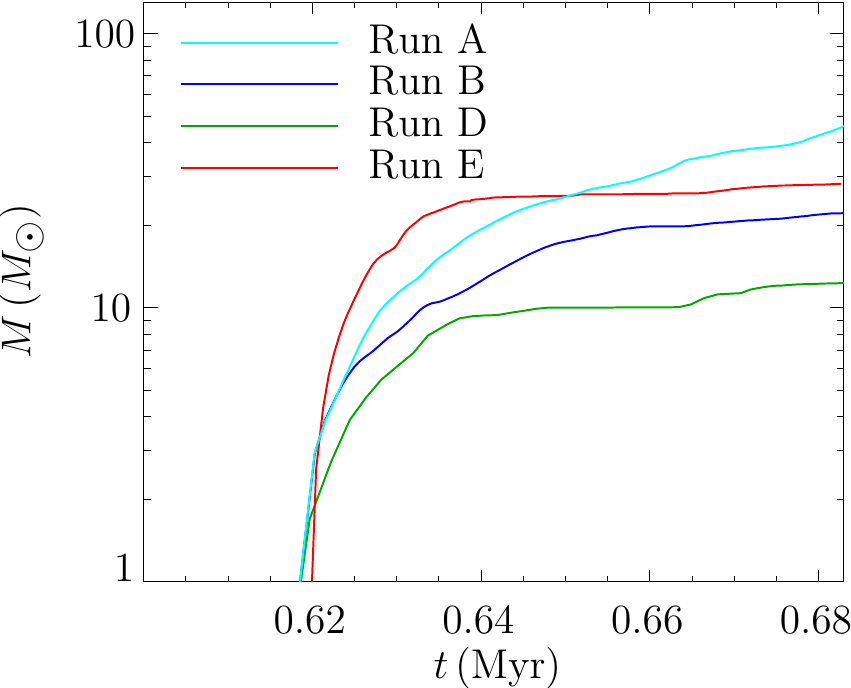}}
\caption{(a) Accretion history for run~E. The first sink that
  forms eventually becomes the only
massive one to form during the simulation runtime. Radiative heating and magnetic support reduce
fragmentation and lead to the formation of only a small number of
sinks in comparison to the equivalent gas dynamical model.
(b) Accretion rates of the first sink particles in run~A, run~B, run~D and run~E. While
the initial accretion phase for the first sink in run~E leads to a higher accretion rate,
the long-term evolution of the accretion rates looks very similar. In contrast to run~B,
however, in which accretion onto the first star stopped when it had reached $23\,M_\odot$
after $0.70\,$Myr, accretion onto the $28\,M_\odot$ star in run~E
still continues (compare (c)).
The accretion rate drops significantly when secondary sink particles form in run~E.
(c) Protostellar masses of the first sink particles in run~A, run~B, run~D and run~E.
Radiative heating (run~B) and presence of magnetic fields (run~E) increase the final masses of the massive stars.
The largest part of the additional mass accretion in run~E compared to run~B is due to the
stronger initial accretion phase.}
\label{fig:accrhist}
\end{figure}

The accretion history for run~E is shown in
Figure~\ref{fig:accrhist}. The first sink particle that forms accretes
more than $28\,M_\odot$ during the simulation runtime of
$0.683\,$Myr, eventually becoming by far the most massive sink.  Only
after a delay of almost $20\,$kyr after the formation of the first sink
do secondary sinks begin to form. None of them gets more massive than
$6\,M_\odot$. In addition to the effect of radiation feedback, which
raises the Jeans mass locally and reduces fragmentation (see
\citealt{petersetal10c} for a thorough discussion), fragmentation is
further reduced in this simulation by the magnetic field, just as is
found in models of low-mass star formation (see Section~1). 
The same effects of reduced fragmentation occur even in our 
model of high-mass star formation with initial $(M/\Phi) \gg
(M/\Phi)_{\mathrm{cr}}$. 

Figure~\ref{fig:accrhist} also shows the accretion rates and
protostellar masses of the first sinks to form in run~A, run~B, run~D, and
run~E for comparison. The accretion rate in the initial accretion
phase is higher in run~E than in the other simulations, but the
long-term accretion behavior does not appear to be substantially
different, apart from a stronger variation beyond
$0.64\,$Myr. However, we note that in run~B, fragmentation-induced
starvation \citep{petersetal10a,petersetal10c} terminates accretion
onto a $23\,M_\odot$ sink, whereas the $28\,M_\odot$ sink in run~E
continues to accrete until the end of the simulation. The central
star in run~E can grow to larger masses because in the initial phase
fragmentation is delayed by magnetic support. This allows the central
object to maintain a high accretion rate for a longer time.
Nevertheless, the accretion rate of the massive star
in run~E drops significantly when secondary sink particles form, since
they form in a dense ring around the central massive sink and to some
extent starve it of material, even though they never cut off accretion
entirely.

The stronger initial accretion phase in run~E compared to
run~B yields the main contribution to the larger final mass of the
massive sink. The magnetic field very efficiently redistributes
angular momentum, resulting in an increased radial mass flux through the
high-density equatorial plane (see Section~\ref{sec:mfs}).  Consequently, the initial accretion
rate in the magnetized simulation (run~E) even lies considerably above
the non-magnetized single sink calculation (run~A).

\begin{figure}
\centerline{\includegraphics[width=8cm]{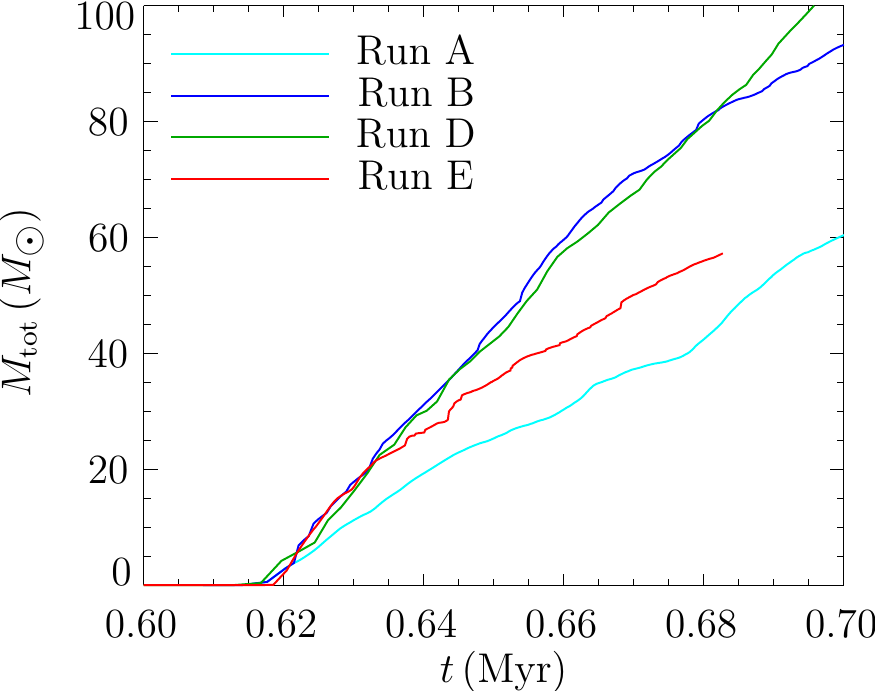}}
\caption{Total accretion history of run~A, run~B, run~D and run~E. While the total accretion rates
of run~B and run~D agree until ionization feedback stops massive star formation in run~B, the
total accretion rate in run~E already starts to decline after $20\,$kyr. The magnetic field in run~E
additionally supports the gas against collapse, reducing the total accretion rate.}
\label{fig:totaccr}
\end{figure}

The total accretion histories of all sink particles combined in each of
the four simulations are contrasted in Figure~\ref{fig:totaccr}.
For the first $20\,$kyr, the total accretion rate of run~B, run~D and run~E is nearly identical.
The accretion rate in these multiple sink simulations is generally higher than in the single sink 
run~A since the large group of sinks accretes from a large volume, without needing to
rely on outward angular momentum transport to deliver material to the direct feeding zone
of the central sink. In the magnetized run~E, no secondary sink particles form during the first $20\,$kyr
(see Figure~\ref{fig:accrhist}), so that the increased accretion rate
in this phase is due to the additional angular momentum transport
performed by the magnetic field.
After the initial $20\,$kyr, the total accretion rate in run~E falls
below run~B and the control run~D, deviating
increasingly with time, but always staying above the accretion rate in run~A.
The accretion histories of run~B and run~D only start to separate at
relatively late time when the ionizing radiation begins to terminate accretion onto the most massive
sinks in run~B (see the discussion in \citealt{petersetal10c}), but the magnetic field additionally
reduces the rate at which gas collapses in run~E. Thus, the total accretion
rate with magnetic fields is lower than without magnetic fields, in
agreement with previous
   work
by \citet{wangetal10} that used a stronger magnetic field.

\subsection{Magnetic Field Structure}
\label{sec:mfs}

As the molecular cloud collapses, it forms a thin, rotationally
flattened structure in the midplane of the rotating cloud. This
disk-like structure is very similar to the one we find in the
run without magnetic fields. As this
structure does not have a Keplerian rotation curve, we avoid calling
it a disk, though it behaves in many ways similar to one. This
structure is essentially what was called a pseudodisk by
\citet{gallishu93a,gallishu93b}.  In the midplane, the dense, rotating gas drags
the magnetic field with it, winding the magnetic field lines into a
toroidal configuration that drives a magnetic tower flow 
\citep{uchishib85,tomisaka98,tomisaka02,matstomi04,machidaetal04,banerjee06b,henfro08}.
However, the magnetic tower can
only build up as long as the field lines are anchored in a coherently
rotating flow.

\begin{figure*}
\centerline{\includegraphics[width=450pt]{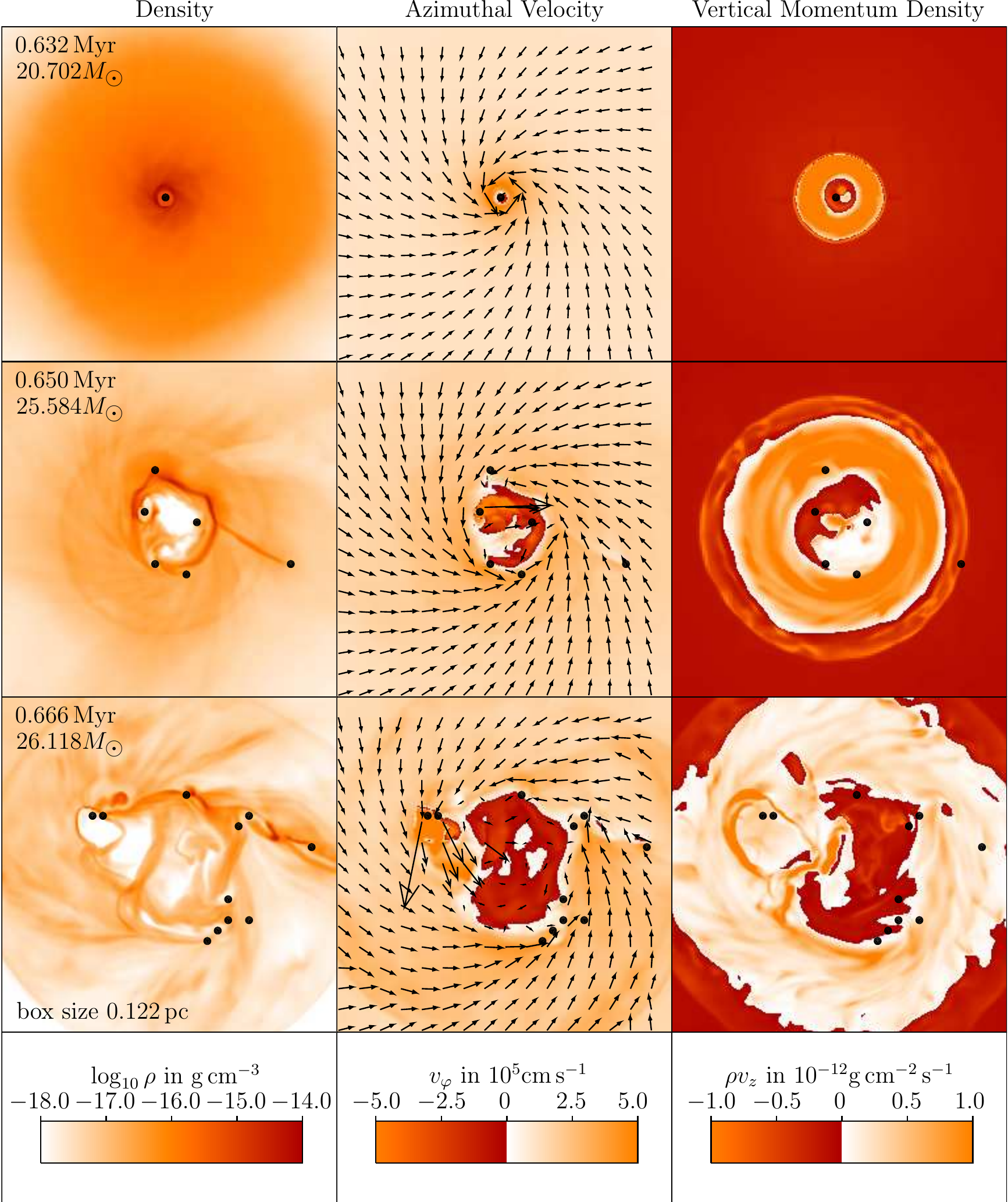}}
\caption{Slices of gas density $\rho$, (left) and azimuthal velocity $v_\varphi$ (middle)
in the midplane and (signed) vertical momentum density $\rho v_z$ (right)
slightly above the midplane
for three different snapshots of run~E. The mass of the most massive star in the cluster
is given in the images. 
   Fragmentation 
sets in after the formation of the first filaments and sink
particles (black dots). The resulting torques can reverse the direction of the azimuthal velocity component.
The arrows indicate the direction of velocity.
Although the cloud is initially rotating counterclockwise (white),
the central region soon begins to rotate clockwise (red), and this region of reversed
azimuthal velocity gradually expands. This has consequences for the driving of the magnetic outflow,
visible in the slices of vertical momentum density (right). The outflow is primarily launched from its outer boundary,
where coherent rotation is maintained, while momentum transport from the inner region is
relatively weak.
The online material contains an animated version of this figure.}
\label{fig:velphi}
\end{figure*}

This velocity coherence is present initially but gets increasingly
disrupted by fragmentation. Figure~\ref{fig:velphi} shows slices of
density, azimuthal velocity and vertical momentum through the midplane at
three different times in run~E. Initially, the whole region is rotating
in the counterclockwise direction. As fragmentation proceeds, regions
with clockwise (negative) azimuthal velocity appear 
near the center of collapse, and these regions expand radially with
time.  This means that the magnetic field lines that are anchored in
the center cannot wind up anymore, and thus the vertical
expansion of the magnetic tower flow stalls. The plots of vertical
momentum of the gas show that the magnetic outflow is thereafter mostly launched
from a ring around the center whose radius grows with
time.
The central regions, where the gas is rotating
clockwise in some regions, eject little further material into the
outflow.

\begin{figure*}
\includegraphics[width=470pt]{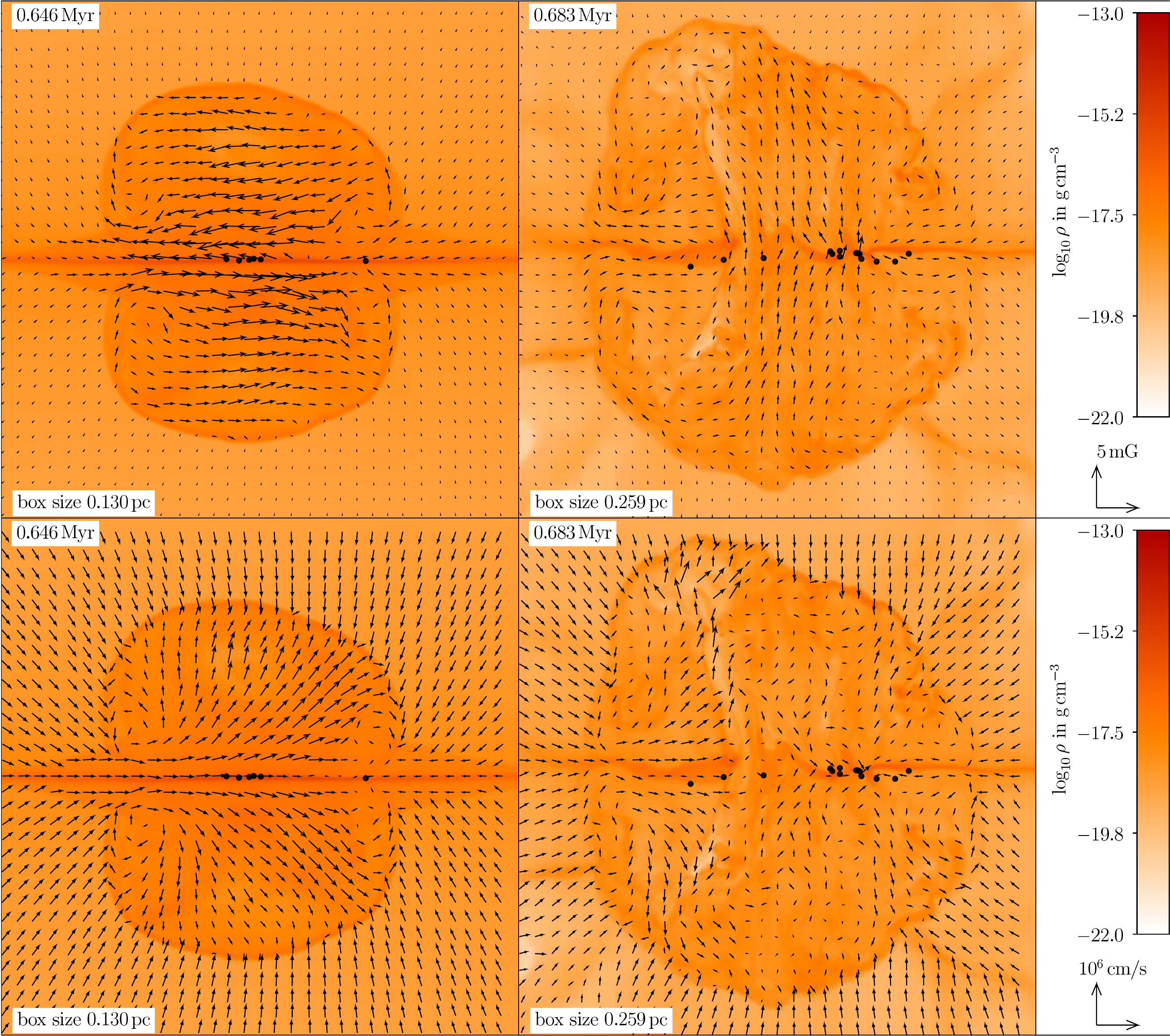}
\caption{Magnetic bubble and field structure. The images show density
  slices parallel to the rotation axis and perpendicular to the midplane through the magnetic bubble.
The arrows indicate the direction of the magnetic field lines (top) and
velocity (bottom), and the black dots show the positions of
sink particles. In the initial phase (left), the field lines wind up to build up a magnetic tower
flow. As the fragmentation occurs, the velocity coherence in the
central regions disappears, so that magnetic field lines
anchored there stop winding up
(right). Hence, the magnetic bubble grows primarily laterally but does not form a more collimated outflow.
Note the different spatial scales in the images.}
\label{fig:bubble}
\end{figure*}

Figure~\ref{fig:bubble} illustrates why gravitational fragmentation
poses a substantial problem to maintaining the magnetic tower
flow. The figure shows density slices perpendicular to the midplane
through the magnetic tower together with the magnetic field. Shortly
after gravitational collapse has formed a dense structure in the midplane and
sink particles have formed, the field lines are nicely aligned and anchored in the
rotating, dense gas, so that a magnetic tower starts to build
up. However, as gravitational fragmentation proceeds, the circular
motion in the inner region gets increasingly disturbed by
gravitational instability as well as angular momentum transport by the
tower flow, resulting in a loss of velocity coherence. Hence, the
magnetic tower grows laterally into regions where the velocity field
remains coherent as fast as it grows in height, so it is unable to
produce a collimated outflow. This structure is a magnetic bubble rather
than the outflow we are familiar with in low-mass star formation
\citep{caband91,bachiller96,reibal01,baletal07}.

Although the velocity field is not turbulent initially, a substantial amount
of turbulence develops during the simulation runtime 
   because of
gravitational fragmentation.
   Conversely, any initial turbulence will die away in a
   sound-crossing time \citep{maclow98}.
We
   do 
suspect that 
   initialization with 
highly supersonic turbulence 
  might
modify the large-scale
dynamics and the shape of filaments in our flattened structure
  but not that it will qualitatively change our results. 

While it has been shown that magnetically driven outflows from low-mass stars
can still form in a turbulent environment \citep{matsuhana10},
it is less clear whether highly collimated, magnetically-driven outflows
will persist around massive stars since turbulent
fragmentation may 
  even enhance the destructive effect of   
gravitational fragmentation
discussed here on the velocity structure of disks around high-mass stars.

\begin{figure*}
\includegraphics[width=470pt]{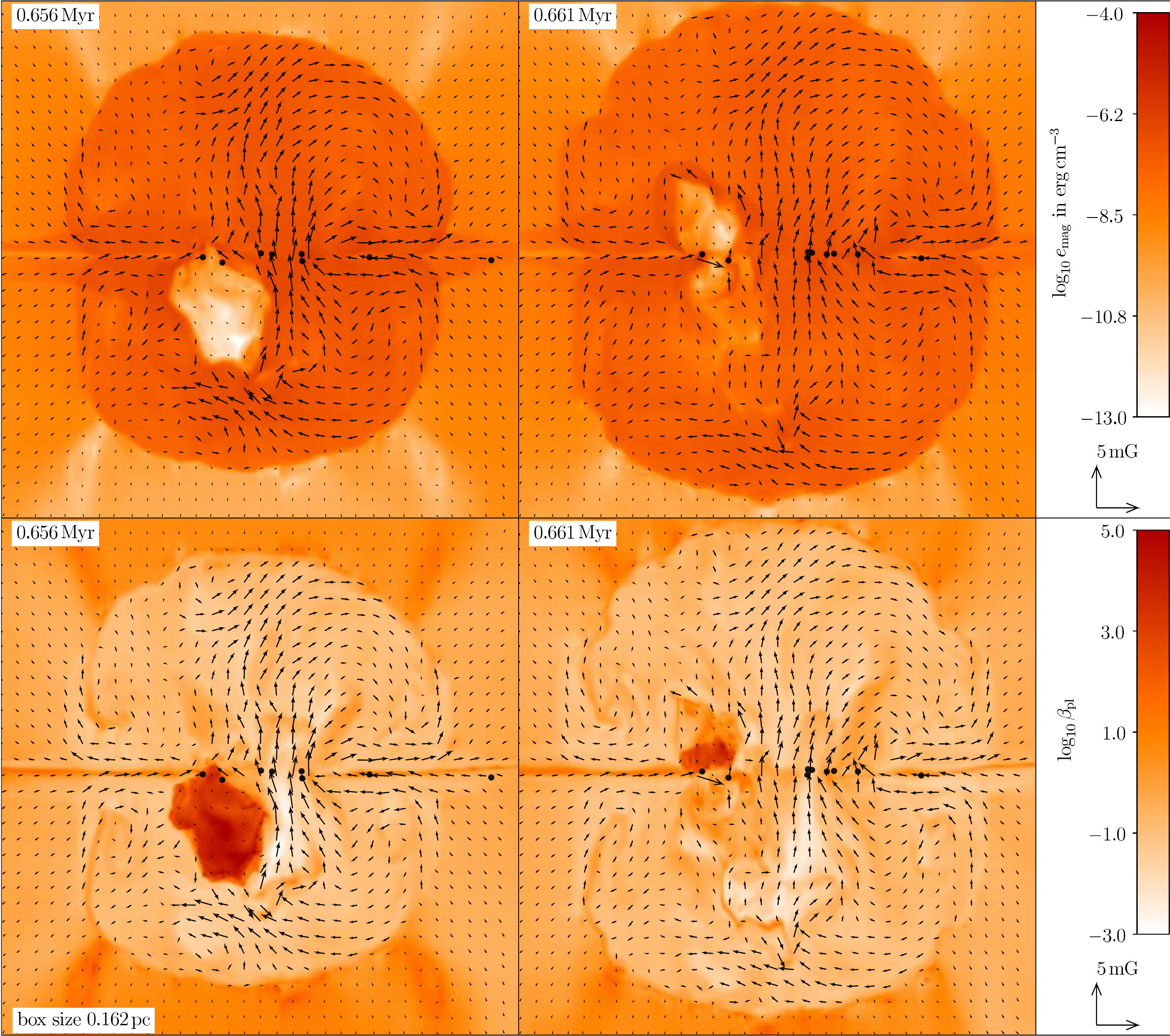}
\caption{Influence of ionizing radiation on magnetic field structure. The images show slices of magnetic
energy density $e_\mathrm{mag}$ (top) and the plasma beta $\beta_\mathrm{pl} = p_\mathrm{th} / p_\mathrm{mag}$
(bottom) along with magnetic field vectors and sink particles. The ionizing radiation emitted by the most
massive star of the cluster creates an \hii\ region, which dramatically decreases the magnetic energy
content in this region and destroys the coherence of the magnetic field structure (left). After
complete recombination of this \hii\ region, the magnetic energy density remains reduced and the magnetic field
structure is not reconstructed (right). The black dots represent sink particles.
The online material contains an animated version of this figure.}
\label{fig:ion}
\end{figure*}

\hii\ region formation further influences the magnetic field structure.
Figure~\ref{fig:ion} shows slices of magnetic energy density and the plasma beta
($\beta_\mathrm{pl} = p_\mathrm{th} / p_\mathrm{mag}$ with the thermal and magnetic pressure $p_\mathrm{th}$
and $p_\mathrm{mag}$, respectively), along with magnetic field vectors. The left-hand
plot demonstrates that the expansion of the \hii\ region driven by
ionization heating reduces the magnetic energy content inside the \hii\ region
of the most massive star in run~E by more than four orders of
magnitude. 
This is because the small amount of flux threading the initially
ionized gas gets redistributed over the large volume that it expands
into because of flux freezing.
The alignment of the magnetic field, which remains visible outside the \hii\ region, also
gets destroyed.  
The right-hand plot shows the same region after the
former \hii\ region has completely recombined. The magnetic field
energy has only increased slightly, since the gas has not been
compressed much,
and the magnetic field orientation is still disarranged. 

\subsection{Gravitational and Magnetic Torques}
\label{sec:torques}

\begin{figure*}
\includegraphics{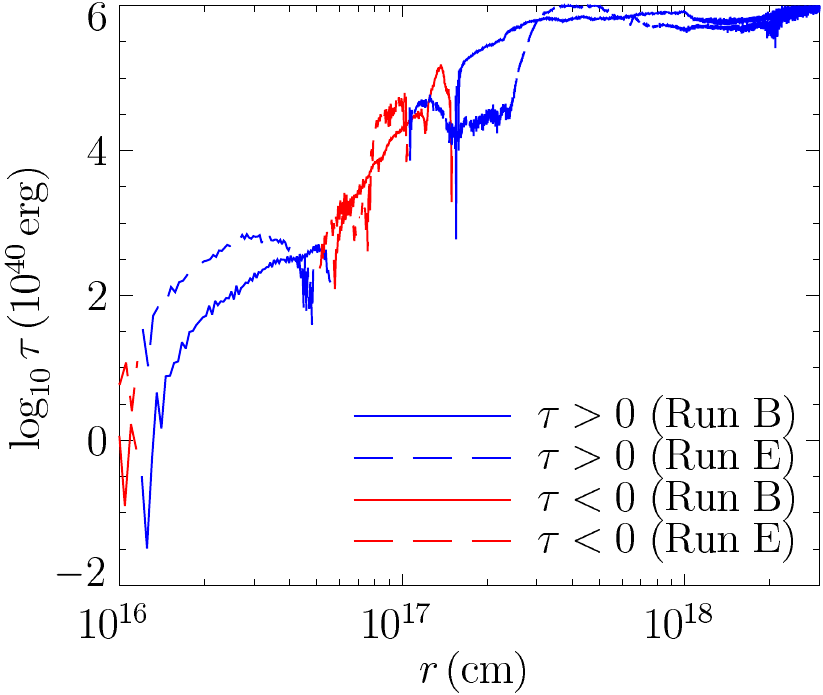}
\includegraphics{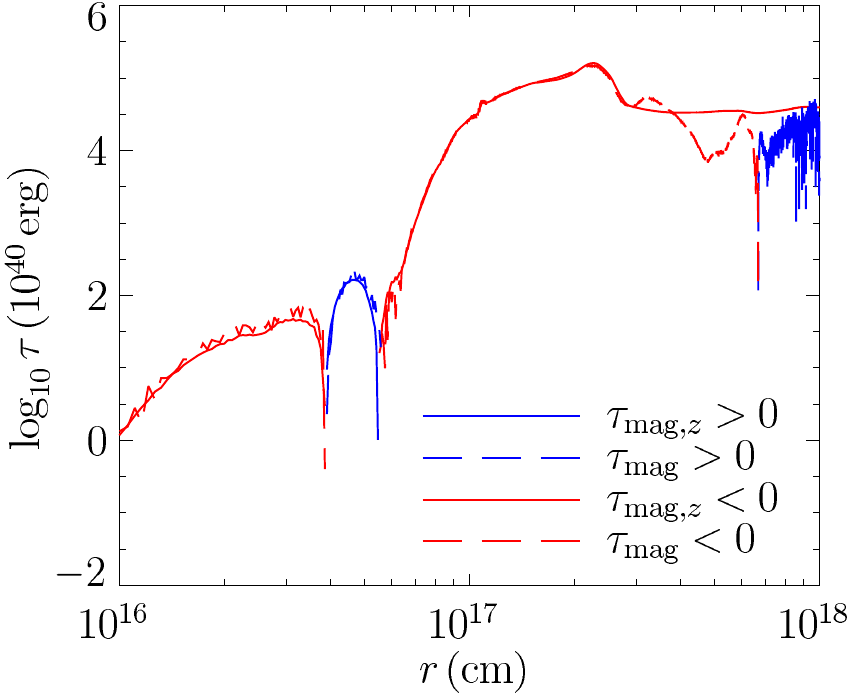}
\caption{(Left) The cumulative torque $\tau$ is plotted for the hydrodynamical simulation
(run~B) and for the MHD simulation (run~E). Negative torques, indicating braking and reversal of the azimuthal
velocity field, appear in both cases. (Right) Cumulative magnetic torques calculated in two different
ways, the $z$-component of Equation~\eqref{eq:taumaggen} and via Equation~\eqref{eq:taumag}.
Deviations become visible only for radii $> 10^{17}\,$cm, when the averaging volumes
become disparate. The snapshot is taken from the end of run~E, where the rotationally flattened flow is strongly fragmented
and asymmetric.}
\label{fig:hdmhdtor}
\end{figure*}

To better understand the reasons for 
   the breakdown  of coherent rotation
and the onset of counter-rotation in the flattened accretion flow,
which ultimately disrupts the magnetic outflow, we study the different
torques
   acting on the gas in the central regions.
The torque $\tau$ exerted on the gas in the midplane \citep[e.g.,][]{banerjee06b} is given by 
\begin{equation}
\tau = r^2 \int v_r \rho R v_\varphi\,\mathrm{d}\Omega
\end{equation}
with the spherical radius $r$, the radial velocity $v_r$, the cylindrical radius $R$,
the azimuthal velocity $v_\varphi$ and the differential solid angle $\mathrm{d}\Omega$. Since the purely hydrodynamical simulation
(run~B) and the MHD calculation (run~E) show fragmentation in the midplane, the negative
torques required to reverse the direction of rotation should be present in both cases. Figure~\ref{fig:hdmhdtor} demonstrates
that his is indeed the case. The braking and reversal of the coherent velocity field
is thus not a new phenomenon, but there are some notable differences
in the fragmentation behavior.

While the only torques acting in the hydrodynamical case are gravitational,
the magnetic field can exert additional torques
\begin{equation}
\bm{\tau}_\mathrm{mag} = \frac{1}{4 \pi} \int \mathbf{r} \times \big[(\bm{\nabla} \times \mathbf{B}) \times \mathbf{B}\big]
\,\mathrm{d}V
\label{eq:taumaggen}
\end{equation}
on the gas in the MHD case. Under the assumption of cylindrical symmetry, the torque
exerted by the magnetic field is
\begin{equation}
\tau_\mathrm{mag} = \frac{1}{4 \pi} r^2 \int B_r R B_\varphi\,\mathrm{d}\Omega
\label{eq:taumag}
\end{equation}
with the radial magnetic field component $B_r$ and the toroidal component $B_\varphi$.
The difference
\begin{equation}
\tau_\mathrm{grav} = \tau - \tau_\mathrm{mag}
\end{equation}
can be attributed to gravitational torques. In principle, both $\tau_\mathrm{mag}$
and $\tau_\mathrm{grav}$ can become negative and force the velocity field to reverse.

Since fragmentation breaks cylindrical symmetry, it is questionable whether
Equation~\eqref{eq:taumag} can still be applied to calculate the magnetic torque.
To test this, we have compared $\tau_\mathrm{mag}$ computed via Equation~\eqref{eq:taumag}
with $\tau_{\mathrm{mag},z}$, the $z$-component of $\bm{\tau}_\mathrm{mag}$ from Equation~\eqref{eq:taumaggen}.
In the latter case, we have evaluated $\tau_{\mathrm{mag},z}$ on a homogeneous grid with a $98\,$AU cell size,
corresponding to the highest refinement level in the adaptive mesh, of dimensions
$0.49 \times 0.49 \times 0.06\,$pc$^3$. A typical snapshot is shown in Figure~\ref{fig:hdmhdtor}.
The agreement is surprisingly good. Although the data is not strictly cylindrically symmetric, the
curves agree very well after the angular averaging process, at least for radii smaller than
$\sim 10^{17}\,$cm, where the regions over which the data is averaged in the two calculations
differ little.
Hence, Equation~\eqref{eq:taumag} is still applicable to study the radial dependence of $\tau_\mathrm{mag}$.

\begin{figure*}
\includegraphics[width=450pt]{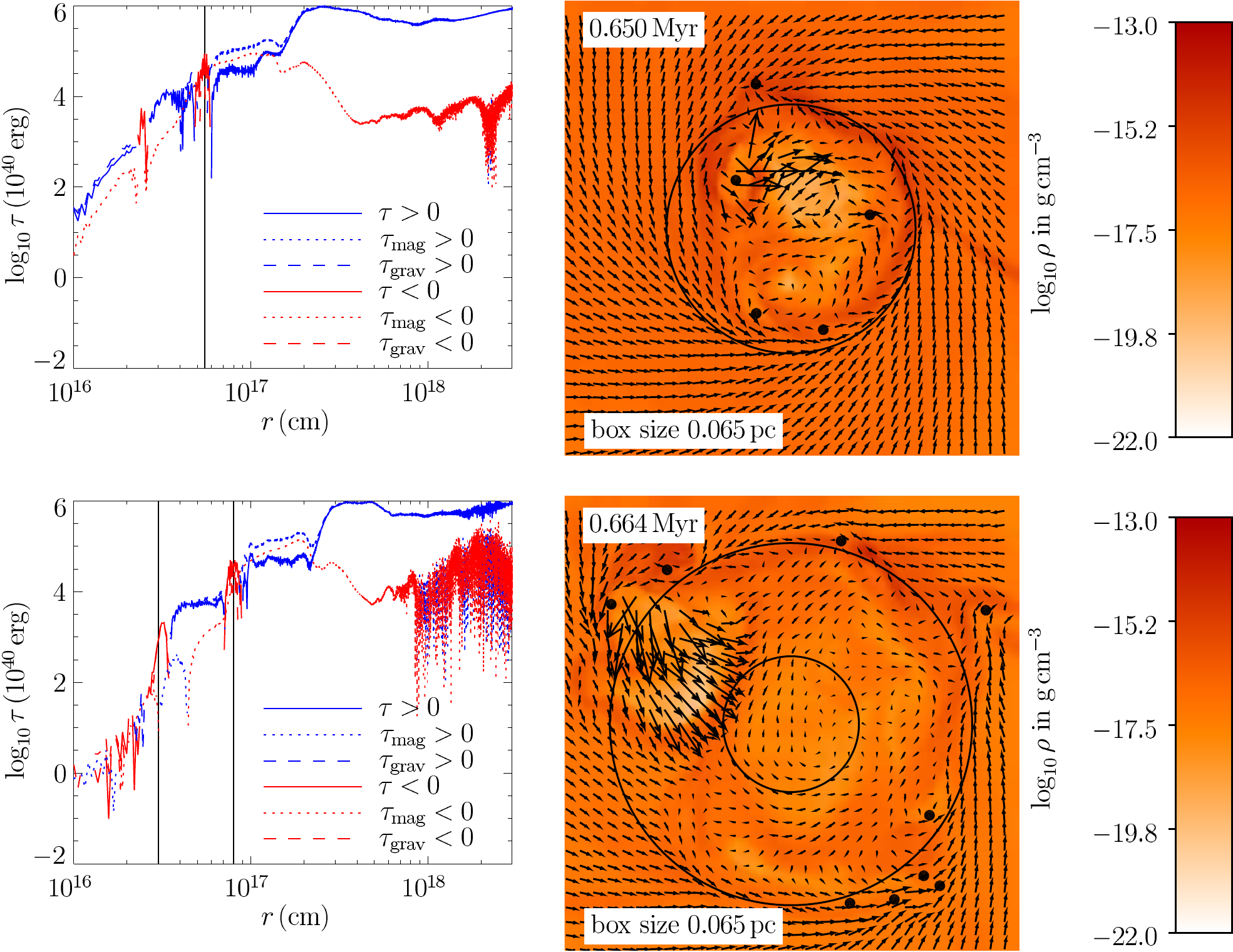}
\caption{Torques and density structure. The figure shows cumulative total, magnetic and
gravitational torques as function of radius and the gas density $\rho$ in the midplane
for two representative snapshots. The magnetic $\tau_\mathrm{mag}$ and gravitational
torques $\tau_\mathrm{grav}$ can reach equipartition locally, but generally the
gravitational torque is considerably larger in magnitude. Regions of negative
total torques (marked as lines) can be associated with circularly arranged
filaments (marked as circles) at the same radii. The black dots
represent sink particles. The arrows indicate the direction of the velocity field.}
\label{fig:torqpanel}
\end{figure*}

The shape of the filaments is substantially different in run~B and run~E.
While the filaments are randomly oriented in the purely hydrodynamical case (run~B),
the magnetic field organizes the filaments into ring-like structures in run~E, as can be nicely
seen in the middle panels of Figure~\ref{fig:velphi}. This raises the question
whether the negative magnetic torques dominate over the gravitational torques in
the MHD simulation. This is generally not the case, as is shown in Figure~\ref{fig:torqpanel}.
Although there are regions where magnetic and gravitational torques are close to equipartition,
the gravitational torques are generally larger in magnitude than the magnetic ones.
Hence, it is commonly the gravitational torques that are responsible for braking
and velocity reversals, in agreement with the observation that we find these
effects also without any magnetic fields in run~B. Magnetic braking appears to
be globally less relevant compared to gravitational torques. However, for the
dynamics in the inner region and the accretion
onto individual sink particles from a smaller-scale accretion
disk, local magnetic braking can nevertheless be important.

Figure~\ref{fig:torqpanel} also shows that the negative total torques can be associated
with circularly arranged filaments in the midplane (marked as circles). Since the filaments often represent
the boundary between clockwise and counter-clockwise rotation (compare Figure~\ref{fig:velphi}),
this is exactly where negative torques are expected. 
When the filaments are circularly
arranged, the angular averaging can yield pronounced negative peaks in the radial
torque distribution. Of course, this special condition is not the only one that can lead
to negative torques, but it regularly occurs.

The velocity vectors plotted in Figure~\ref{fig:torqpanel} indicate that the negative
torques 
     cause a
reversal of the radial velocity component at the filaments. 
     This 
reversal of the radial velocity component is not simply the effect
of an expanding motion near the center, but a result of 
    the gravitational torques exerted by the non-uniform gravitational
    instability in the disk.

\subsection{Gravitational Instability and Fragmentation}

The fragmentation of the rotationally flattened accretion flow leads to the formation
of relatively long-lived, circularly arranged filaments in run~E. This is in contrast
to the situation in run~B, where the filaments are generally less elongated
and their shape and location cannot directly be associated with the rotating
flow (see Figure~\ref{fig:diskbetatoomre}). To understand the 
  apparent
stability of this structure, we investigate the relative importance of two
contributing mechanisms: 
support by magnetic pressure (Section~~\ref{sec:magpres})
and 
support by 
strong shear across the filaments (Section~\ref{sec:shear}).

\subsubsection{Support by Magnetic Pressure}
\label{sec:magpres}

\begin{figure*}
\centerline{\includegraphics{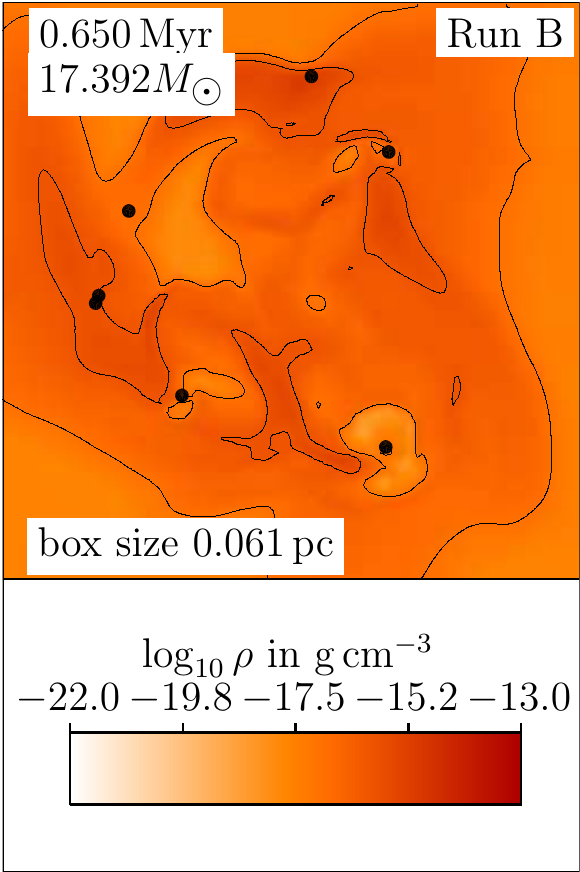}
\includegraphics{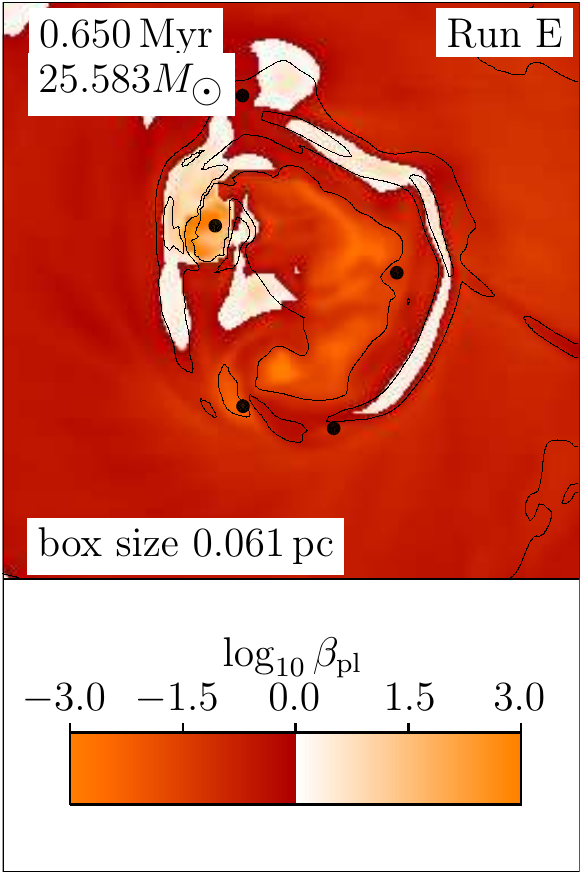}
\includegraphics{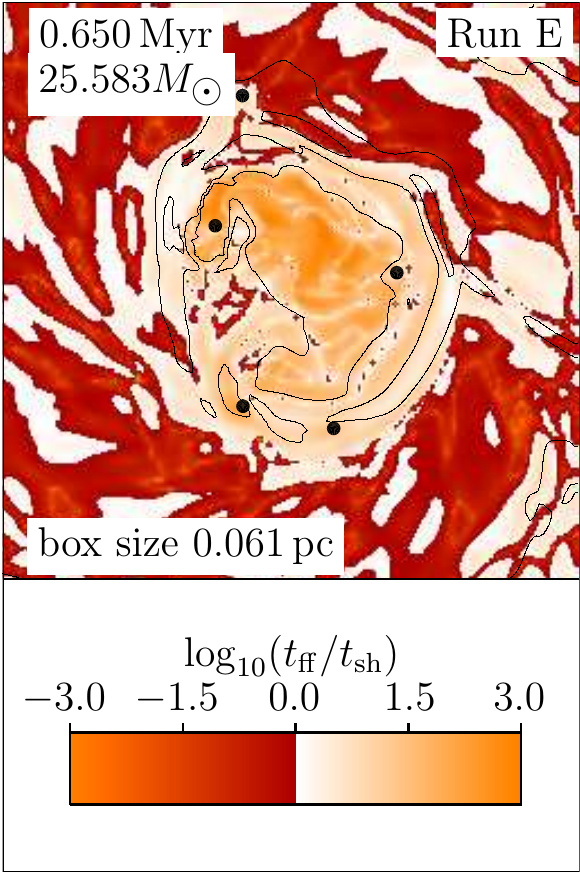}}
\caption{(Left) Slice of gas density in the hydrodynamical simulation run~B. 
The filaments are much more roundish and are not stretched between rotating and counter-rotating
gas. The black dots represent sink particles. The contours show the density structure.
(Middle) The plasma beta $\beta_\mathrm{pl} = p_\mathrm{th} / p_\mathrm{mag}$ in the midplane of run~E.
Outside the \hii\ region around the $26\,M_\odot$ star, the magnetic pressure is generally larger
than the thermal pressure and hence is dynamically relevant. However, $p_\mathrm{th}$ is slightly
larger than $p_\mathrm{mag}$ in the filaments, and thus magnetic pressure plays no dominant role
in stabilizing them.
(Right) The ratio $t_\mathrm{ff}/t_\mathrm{sh}$ for the same region
from run~E. The filaments, marked with the density contours, are on the boundary between
the stable and the unstable regime.}
\label{fig:diskbetatoomre}
\end{figure*}

Though the magnetic field appears to be not the main driver of negative
torques, it is clearly dynamically relevant for the gas dynamics within
the accretion flow. This is demonstrated by examining the plasma beta
$\beta_\mathrm{pl} = p_\mathrm{th} / p_\mathrm{mag}$ in the midplane, as shown
in Figure~\ref{fig:diskbetatoomre}. Outside of the \hii\ region around the massive
star, the magnetic pressure $p_\mathrm{mag}$ exceeds the thermal pressure
$p_\mathrm{th}$ generally, except in the dense filaments. In the central region,
$p_\mathrm{mag}$ is more than two orders of magnitude larger than $p_\mathrm{th}$,
but near the filaments $p_\mathrm{mag}$ and $p_\mathrm{th}$ reach
   equality.

The plasma beta $\beta_\mathrm{pl}$ also quantifies the importance of magnetic support
against gravitational collapse in a linear perturbation analysis.
The Toomre $Q$-parameter is defined as
\begin{equation}
Q = \frac{\kappa c_\mathrm{s}}{\pi \Sigma G}
\end{equation}
with epicyclic frequency $\kappa$, sound speed $c_\mathrm{s}$, surface density $\Sigma$, and
Newton's gravitational constant $G$. Linear gravitational instability sets in when $Q < 1$
\citep{toomre64,goldlynd65}. When magnetic fields are taken into account \citep{kimost01},
one can define the magnetic Toomre parameter 
\begin{equation}
Q_\mathrm{M} = \frac{\kappa \left(c_\mathrm{s}^2 + v_\mathrm{A}^2\right)^{1/2}}{\pi \Sigma G}
\end{equation}
with the Alfv\'en velocity $v_\mathrm{A}$. Since $\beta_\mathrm{pl} = c_\mathrm{s}^2 / v_\mathrm{A}^2$,
the ratio of $Q$ and $Q_\mathrm{M}$ is
\begin{equation}
\frac{Q_\mathrm{M}}{Q} = \sqrt{1 + \frac{1}{\beta_\mathrm{pl}}}.
\end{equation}
Because $\beta_\mathrm{pl}$ is larger than unity in the filaments, $Q_\mathrm{M}$ and $Q$
only differ by a factor of order unity there,
so that magnetic pressure plays no crucial role in stabilizing the filaments.

\subsubsection{Support by Shearing Motion}
\label{sec:shear}

Since the Toomre analysis as presented above only applies to linear perturbations
to Keplerian rotation, it entirely neglects the strong shear along the filaments
(compare Figure~\ref{fig:velphi}). Because the gas on one side of a filament
is moving clockwise while the gas on the other side is moving counter-clockwise,
there is a large velocity gradient across the filament. This velocity gradient
appears able to stabilize the filament. To see how big this effect is,
we define the shear timescale
\begin{equation}
t_\mathrm{sh} = \frac{1}{|\bm{\nabla} \times \mathbf{v}|}
\end{equation}
with the vorticity $\bm{\nabla} \times \mathbf{v}$ and compare this with the free-fall
time scale
\begin{equation}
t_\mathrm{ff} = \sqrt{\frac{3 \pi}{32 G \rho}}.
\end{equation}
The ratio $t_\mathrm{ff}/t_\mathrm{sh}$ is plotted in Figure~\ref{fig:diskbetatoomre}.
The filaments lie in the marginally stable
regime, so that this
simple timescale comparison is not totally conclusive. There is, however, a clear
difference between the organization of the filaments in run~B and run~E visible in Figure~\ref{fig:diskbetatoomre}.
In run~E, the filaments get stretched by the shear and organize circularly, while
the filaments in run~B are much more roundish in shape. In run~E, the filaments
separate rotating and counter-rotating gas, which is not generally the case in run~B.
Since it is the magnetic field that organizes the filaments in this way, the magnetic
field indirectly stabilizes the filaments, though not directly through magnetic pressure.

\subsection{Relevance for Protostellar Jets}

These findings are 
    immediately applicable to 
the large-scale rotating, infalling flow,
far away from the actual radius where high-velocity protostellar jets
    must be
launched.
Since the rotation of the launching region
determines the outflow velocity, our larger-scale magnetic tower flow,
with peak velocities around
     5~km~s$^{-1}$,  (compare
Figure~\ref{fig:bubble})
is slower by a factor of more than 20 than optical jets with
velocities 
    exceeding
$100\,$km\,s$^{-1}$ \citep[eg.][]{baletal07}.

However, the behavior seen here may have important implications for understanding the role
of jets in high-mass star formation.  Such jets
are driven from radii of less than an astronomical
unit in the accretion disks of low mass
stars.  We do not resolve the accretion flow to these small radii in
our models.  However, the huge accretion rates required for high mass
star formation suggests that those inner regions must also be
highly perturbed and may likely be gravitationally unstable,
and thus susceptible to the same
fragmentation and destruction of velocity coherence
that we have seen
in our models. Observations of a counter-rotating accretion disk
\citep{remiholl06} demonstrate that this possibility exists.
Simplified one-dimensional calculations \citep{vaidyaetal09}
indicate that the temperature may grow strongly enough below a
radius of 1\,AU to stabilize the disk, but it may be difficult for
the magnetic tower to build up on such small scales if the magnetic
field outside this radius is totally disarranged.  
   Models of dynamo generation of fields in such disks including non-ideal MHD effects are
   required to understand whether coherent fields capable of driving
   jets can form on such small scales.
Our findings thus 
   raise the question whether
highly collimated, magnetically-driven jets from 
  massive
protostars
  can 
survive.
A negative answer could also explain why outflows from massive stars appear to
be less collimated \citep{arceetal07,beuthshep05}:
they are driven by the ionizing radiation
instead of coherently rotating magnetic fields.
Indeed, \citet{petersetal10a} presented a detailed comparison
between the ionization-driven outflows in the purely hydrodynamical
simulation with observations of W51e2 and found that the 
molecular and ionized parts of the outflow have a very similar
velocity structure, both qualitatively in the characteristic
features as well as quantitatively in the magnitude of the velocity
gradient.

\subsection{Morphology of \hii\ Regions}

\begin{figure*}
\centerline{\includegraphics[width=450pt]{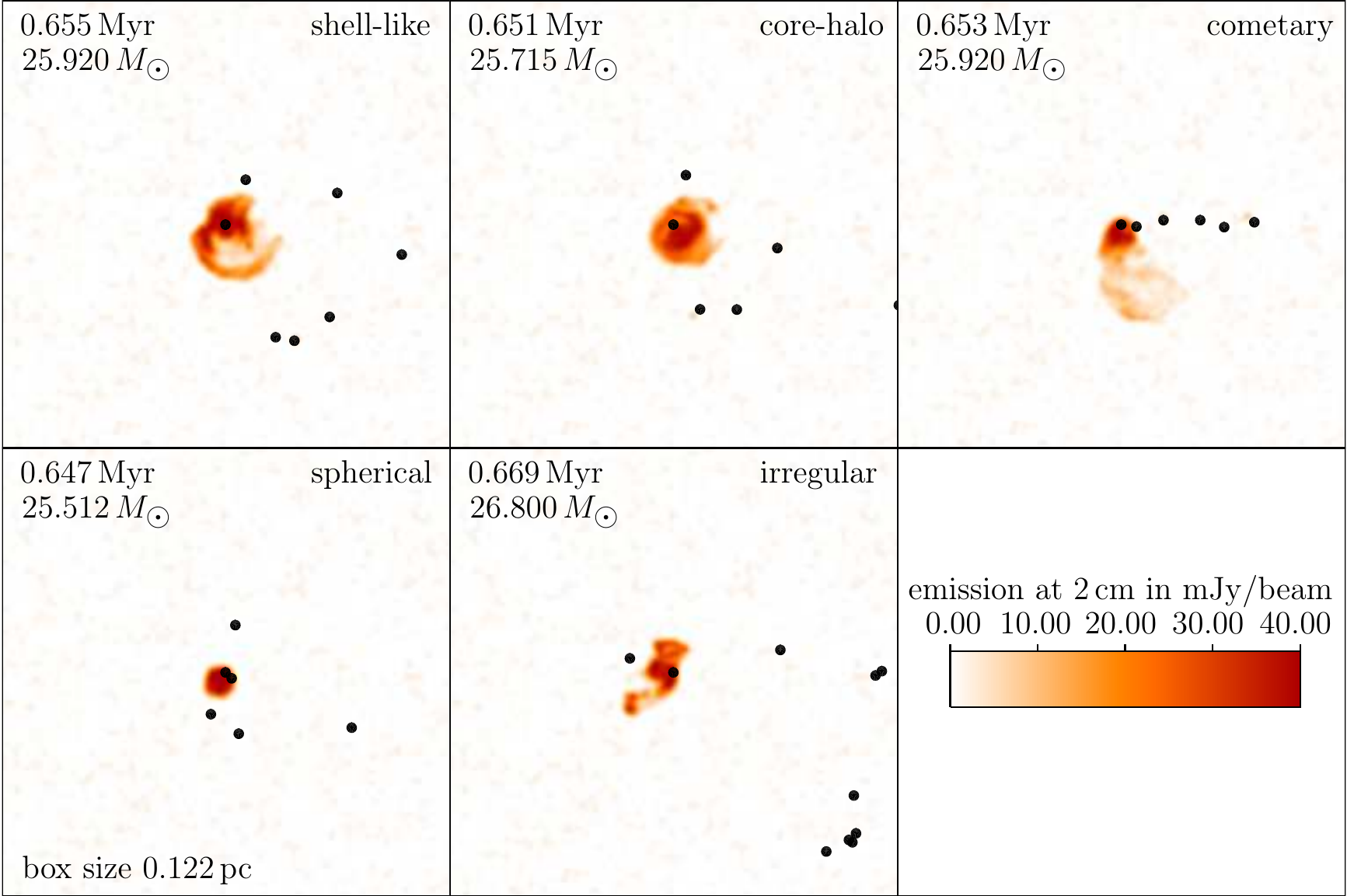}}
\caption{\hii\ region morpholgies in run~E. The figure shows synthetic maps of
free-free emission from ultracompact
\hii\ regions around the massive protostar at different time steps and from
different viewing angles. The cluster is assumed to be $2.65\,$kpc away, the full width
at half maximum of the beam is $0\farcs14$ and the noise level is $10^{-3}$Jy.
This corresponds to typical VLA parameters at a wavelength of $2\,$cm.
The protostellar mass of the central star which powers the \hii\ region is
given in the images. Black dots represent sink particles.}
\label{fig:morph}
\end{figure*}

To study the morphological types of the \hii\ regions in run~E in more
detail, we have generated synthetic VLA observations at $2\,$cm,
following the algorithm described earlier \citep{petersetal10a,petersetal10b}.
Figure~\ref{fig:morph}
shows \hii\ regions of all morphological types found in run~E. There
seems to be no systematic difference between the morphologies
in the magnetic and purely hydrodynamic simulations \citep{petersetal10a,petersetal10b}.

\begin{figure*}
\includegraphics[width=470pt]{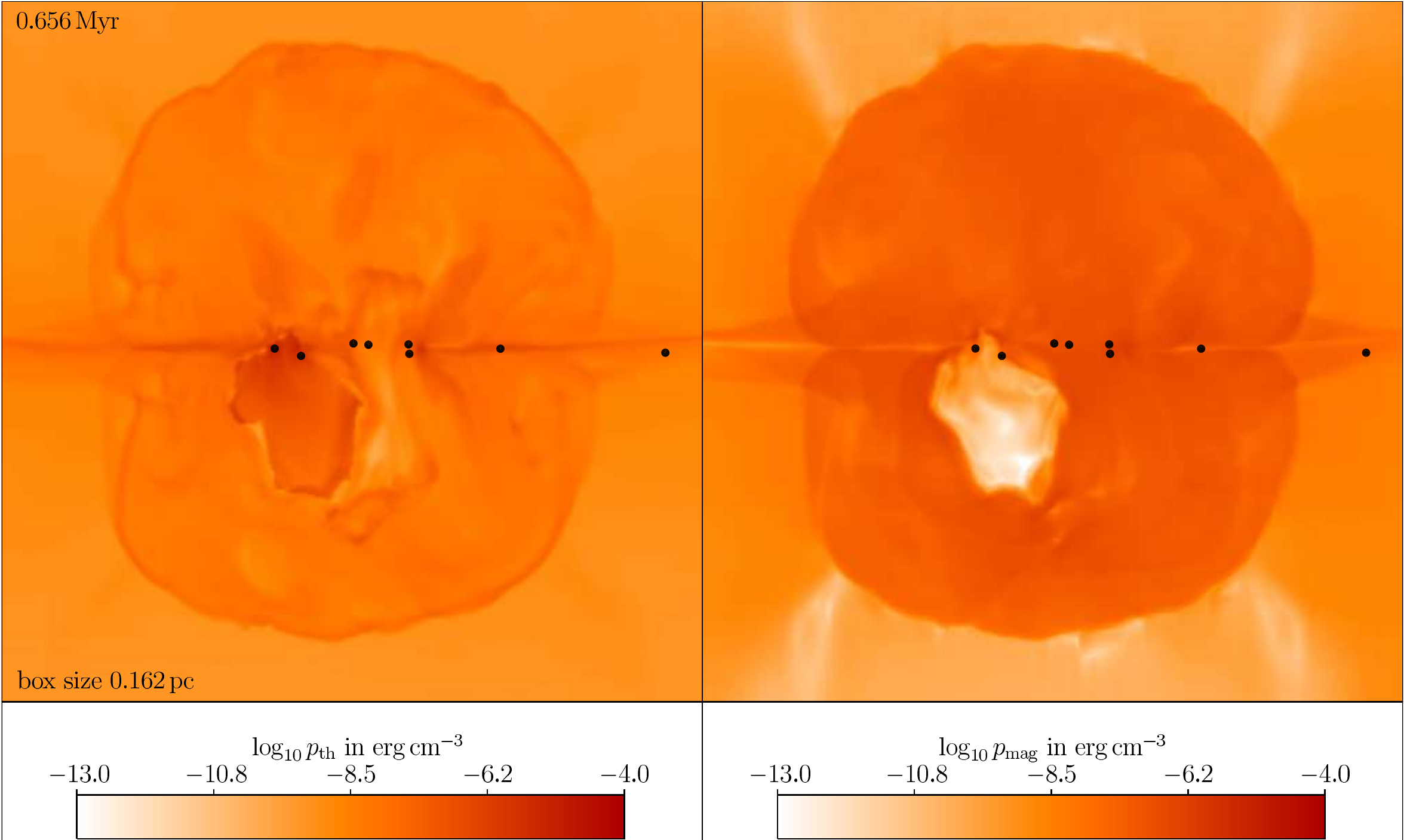}
\caption{Comparison of thermal and magnetic pressure for the data from the
left-hand panels in Figure~\ref{fig:ion}. The thermal pressure $p_\mathrm{th}$ inside
the \hii\ region (left) is of comparable magnitude 
     to
the magnetic pressure $p_\mathrm{mag}$
outside the \hii\ region (right). Thus, magnetic pressure plays a significant
role in constraining the size of expanding \hii\ regions.
The black dots represent sink particles.}
\label{fig:pressure}
\end{figure*}

Although the magnetic field does not appear to have any influence
on the \hii\ region morphologies themselves, it does play a significant
role in determining their size. Figure~\ref{fig:pressure} shows
a comparison of thermal and magnetic pressure for the \hii\ region from
Figure~\ref{fig:ion}. The thermal pressure inside the \hii\ region
is of the same order as the magnetic pressure immediately outside
the \hii\ region. Thus, the magnetic pressure yields an important contribution to the
total pressure in the environment of the \hii\ region, constraining
its expansion. We therefore expect \hii\ regions in the presence
of strong magnetic fields to be generally smaller than without the
magnetic field.

\section{Summary and Conclusion}
\label{sec:summary}

We have presented the first three-dimensional,
radiation-magnetohydrodynamical collapse simulation of massive star
formation including heating by both ionizing and non-ionizing
radiation. We used sink particles to represent accreting
protostars. We compared the action of the magnetic field in this
high-mass star formation simulation to the major ways that it acts in
low-mass star formation.  As we summarized in the introduction, these
include giving support against gravitational collapse, transferring
angular momentum outward, and driving outflows.

We find that, although we started with a mass-to-flux ratio that was more than an
order of magnitude supercritical, $(M/\Phi) = 14
(M/\Phi)_{\mathrm{cr}}$, the magnetic field still can reduce
fragmentation (Section~\ref{sec:accrhist}), just as it does in low-mass
star formation. The additional magnetic support prevents the
gas from collapsing into as many secondary fragments compared to the non-magnetic case, leaving
the most massive protostar with a larger gas reservoir. For the same reason, the total accretion
rate is reduced.
However, it is important to note that the magnetic field
    does not
suppress fragmentation completely. The number of fragments is reduced
at most by a factor of two. Similar findings are reported by
Hennebelle et al. (2010, A\&A, submitted).

Additionally, magnetic braking near the center reduces the angular momentum of the
collapsing gas even further.  Thus, more material is transported radially inwards
onto the massive central protostar. This is important since the
distribution of mass accretion among the protostars in the
fragmentation-induced starvation scenario depends sensitively on the
(in)efficiency of gas transport towards the center by
internal torques \citep{petersetal10a,petersetal10c}.

The combination of these two effects causes the formation of a 50\%
more massive central star compared to our equivalent simulation
without magnetic fields (Figure~\ref{fig:accrhist}c). If the magnetic field strengths
inferred by \citet{falgaroneetal08} are typical 
   throughout high-mass clumps, and not just in their high-density
cores, magnetic
braking and reduction of fragmentation may 
         be even
stronger than reported here.

The winding up of the magnetic field into a toroidal configuration
leads to the formation of a large-scale tower flow,
surrounding all the protostars. 
However, two effects tend to weaken and broaden the outflow.  The first one
is the gravitational fragmentation of the accretion flow. This disrupts
the circular motion necessary to drive the tower flow in the central region. Second, the thermal
pressure of ionized gas is by far larger than the magnetic pressure
and hence dynamically dominant within the \hii\ region. The resulting
expansion of the \hii\ region dramatically reduces the magnetic energy
content within it, and tangles the field lines, further weakening the
flow.  

The torques required to brake the large-scale rotating flow and reverse the azimuthal velocity
are primarily gravitational, though magnetic and gravitational torques
can reach equipartition locally.
Magnetic braking appears to be
of minor importance for the global gas dynamics, but not necessarily for the local
accretion onto protostars.
Again, if stronger magnetic fields are imposed,
magnetic braking 
    might become qualitatively 
more important.
The shape and dynamics of the filaments is vastly
different in the magnetohydrodynamical (run~E) and purely gas dynamical (run~B) case: while
the filaments in run~B are disorganized, roundish and quickly collapse to form
stars, the filaments in run~E are stretched between clockwise and counter-clockwise
rotating gas and can be maintained over a large fraction of the
simulation runtime.

If the fragmentation process observed in our simulations holds down
to the smallest scales of the massive accretion disk, as suggested
by the very high accretion rates of the order $10^{-3}\,M_\odot\,$yr$^{-1}$
inevitably needed to form a massive star, the magnetic field can only
wind up at radii less than 1\,AU,
where the temperature is high enough to suppress gravitationl instability
of the disk \citep{vaidyaetal09}.
In this case, magnetically-driven, steady
jets around massive protostars can only survive until gravitational
fragmentation disrupts uniform rotation, and ionizing radiation becomes
dynamically relevant. We call this process fragmentation-induced
outflow disruption. The fast jets launched from the inner disk region
should then be highly episodic like the accretion rates.
 The uncollimated outflows from massive stars may 
be better explained as driven by the ionization feedback.

\acknowledgements{
We thank the anonymous referee for very useful comments that helped to improve the paper.
T.P. is a Fellow of the Baden-W\"{u}rttemberg Stiftung funded by their program International
Collaboration II (grant P-LS-SPII/18). He also acknowledges support from an Annette Kade Fellowship for his
visit to the American Museum of Natural History and a Visiting Scientist Award of the Smithsonian
Astrophysical Observatory (SAO).
R.S.K.\ acknowledges financial support from the Baden-W\"{u}rttemberg Stiftung
via their program International Collaboration II (grant P-LS-SPII/18) and from the German
Bundesministerium f\"{u}r Bildung und Forschung via the ASTRONET project STAR FORMAT (grant 05A09VHA).
R.S.K. furthermore gives
thanks for subsidies from the Deutsche Forschungsgemeinschaft (DFG) under
grants no.\ KL 1358/1, KL 1358/4, KL 1359/5, KL 1358/10, and KL 1358/11, as well as from a Frontier
grant of Heidelberg University sponsored by the German Excellence Initiative.
M.-M.M.L. was partly supported by NSF grant AST 08-35734.
R.B. is funded by the DFG via the Emmy-Noether grant BA 3706/1-1.
We acknowledge computing time at the Leibniz-Rechenzentrum in Garching (Germany), the NSF-supported
Texas Advanced Computing Center (USA), and at J\"ulich Supercomputing Centre (Germany). The FLASH code
was in part developed by the DOE-supported Alliances Center for Astrophysical Thermonuclear
Flashes (ASCI) at the University of Chicago.}

\end{document}